\input epsf
\documentstyle[twoside,fleqn,epsfig,espcrc2]{article}

\newcommand{\AmS}{{\protect\the\textfont2
  A\kern-.1667em\lower.5ex\hbox{M}\kern-.125emS}}
\title{Unquenched Studies Using the Truncated Determinant Algorithm}
\author{A.~Duncan\address{Dept. of Physics and Astronomy,
University of Pittsburgh, Pittsburgh, PA 15260},%
E.~Eichten\thanks{Presenter}\address{Fermilab, PO Box 500, Batavia, IL 60510}%
and 
H.~Thacker\address{Dept. of Physics, University of Virginia, 
Charlottesville, VA 22901}}
\begin{document}
\begin{abstract}
A truncated determinant algorithm is used to study the physical
effects of the quark eigenmodes associated with eigenvalues below 420 MeV. 
This initial high statistics study focuses on coarse ($6^4$) lattices 
(with O($a^2$) improved gauge action), 
light internal quark masses and large physical volumes.
Three features of full QCD are examined: 
topological charge distributions, 
string breaking as observed in the static energy
and the eta prime mass.
\end{abstract}

\maketitle

\section{TDA Simulations on Coarse Lattices}

In the truncated determinant approach (TDA) to 
full QCD,  the quark determinant, $ {\cal D}(A) = {\rm Det}(H) =
{\rm Det}(\gamma_{5}(D\!\!\!\!/(A)-m))$ is split-up gauge invariantly 
into an infrared part and an ultraviolet part\cite{truncdet}.
\begin{equation}
\label{eq:split}
{\cal D}(A)={\cal D}_{IR}(A){\cal D}_{UV}(A)  
\end{equation}
The ultraviolet part, ${\cal D}_{UV}$, can be accurately fit 
by a linear combination of a small number of Wilson loops\cite{looppaper}.
The infrared part ${\cal D}_{IR}(A)$ is defined as the product of 
the lowest $N_{\lambda}$ positive and negative eigenvalues of $H$, 
with $|\lambda_{i}|\leq \Lambda_{QCD}$. 
The eigenvalues $\lambda_{a}$ of $H$ are gauge invariant 
and measure quark off-shellness.
The cutoff (for the separation in Eq. \ref{eq:split}) 
is tuned to include as much as
possible of the important low-energy chiral physics 
of the unquenched theory while leaving the fluctuations 
of $\ln {\cal D}_{IR}$ of order unity after each sweep updating
all links with the pure gauge action (assuring a tolerable acceptance
rate for the accept/reject stage)\cite{truncdet,stringbreak}. 
This procedure works well even for kappa values 
arbitrarily close to kappa critical. 

Initial studies using TDA focus on the qualitative 
physical effects of the inclusion of the infrared quark eigenmodes.
For this purpose, coarse lattices with large physical volumes 
are appropriate. 
Here we report the study of $6^4$ lattices with an O($a^2$) 
improved gauge action
\begin{eqnarray}
 S_G &=& \beta_{\rm plaq} \sum_{\rm plaq} \frac{1}{3} {\rm Re Tr(1-U_{plaq})}
\nonumber \\ 
      & & + \beta_{\rm trt} \sum_{\rm trt} \frac{1}{3} {\rm Re Tr(1-U_{trt})}
\end{eqnarray}
with $\beta_{\rm plaq}= 3.7$ and $\beta_{\rm trt}/\beta_{\rm plaq} = 1.04$.
This $O(a^2)$ improved gauge action was adjusted (as in Ref. \cite{impgauge}) 
to have the lattice scale $a \approx 0.4$ fm.  

We used naive Wilson fermions with light quark masses 
($\kappa = 0.2050$, $0.2044$ and $0.2038$); and keep
the lowest 840 eigenvalues of ${\cal D}$,  
i.e. $\Lambda_{\rm cut} \approx 420$ MeV ($1/a = .5$ GeV). 
No terms modelling ${\cal D}_{UV}$ were included in the gauge action.

\begin{figure}
\epsfig{figure=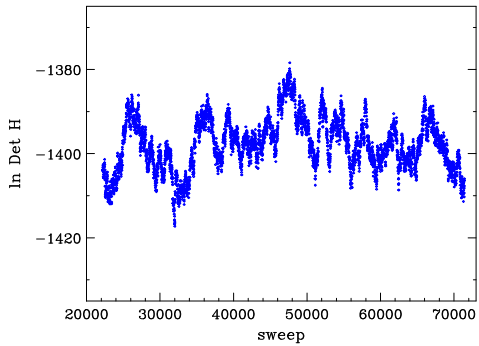, height=0.60\hsize}
\caption{Equilibrated quark determinant in 
TDA simulations on a coarse $6^4$ lattice. ($\kappa =.2050$)
The lowest 840 eigenvalues included in ln~Det~$H$.} 
\label{fig:det1}
\end{figure}

With $\kappa = .2050$,  approximately 15,000 full steps were required 
for the lattice configurations to equilibrate (reflecting critical
slowing down - a few hundred suffice on small physical volumes). 
The equilibrated behaviour of ${\cal D}_{IR}$ is shown for 50,000 sweeps 
in Figure \ref{fig:det1}. 

In the remaining sections we consider some 
physical quantities that differentiate between quenched and full QCD. 

\section{Topolopy}
The topological charge, $Q$, can be expressed 
in terms of the eigenvalues of the Wilson-Dirac operator. 
\begin{equation}
 Q = {1 \over 2\kappa} (1 - {\kappa \over\kappa_{c}}) \sum_{i=1}^{N} {1 \over \lambda_i}
\end{equation}
This sum is quickly saturated by the low eigenvalues.

In full QCD configurations very small eigenvalues of $H$
are suppressed by the quark determinant factor. 
In particular, non-zero topological charges must be suppressed
in the chiral limit ($m_q \rightarrow 0$). 
Furthermore,  the functional dependence of the topological charge distribution, $P_Q$,  
on the light quark mass $m_q$ is predicted by the chiral analysis of
Leutwyler and Smilga\cite{LeutSmil}.  
\begin{equation}
 P_Q = I_Q(x)^2 - I_{Q+1}(x) I_{Q-1}(x) \label{eq:LS} 
\end{equation}
where $x = 1/2 V f^2_{\pi}m^2_{\pi} = V m_q <\overline \psi \psi >$,
$I_{Q}$ are modified Bessel Functions of order Q 
and V is the total space-time volume.

\begin{figure}
\epsfig{figure=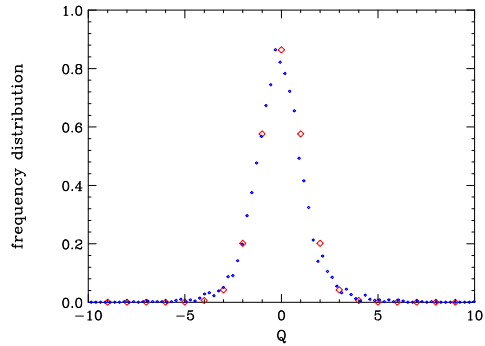, height=0.60\hsize}
\caption{Topological charge distribution for $\kappa =.2050$ on 
a coarse $6^4$ lattice (dots). Expected chiral behaviour (diamonds)
is also shown.}
\label{fig:top1}
\end{figure}

The qualitative agreement with the expected behaviour 
of Eq. \ref{eq:LS} has been reported previously\cite{truncdet}.
Using the value of $x=3.8$ determined from the lattice calculations
($m_{\pi} = .39$ and $f_{\pi} = .20$ for $\kappa = .2050$),
excellent quantitative agreement between theory and lattice calculations 
is shown in Figure \ref{fig:top1}.  
The data from a single kappa value (e.g. $\kappa = .2050$) 
determines $\kappa_c$.  Equating the calculated and predicted
behaviour we find the factor $1-{\kappa/\kappa_{c}} = 0.0082$, i.e.
$\kappa_{c} = .2067$. 

\section{Static Energy}

The static energy $V(\vec{R})$ is defined as:
\begin{equation}
  V(\vec{R}) \equiv \lim_{T\rightarrow\infty}\ln{\frac{W(\vec{R},T-1)}{W(\vec{R},T)}}
\end{equation} 
In Coulomb gauge, $W(\vec{R},T)$ is the Wilson line correlator
and $V(R)$ is the energy of the lowest physical state coupling the
static quark-antiquark state (with separation $R$) to the vacuum.
As $T\rightarrow \infty$, V(R) is approached from above.

Light quark masses implies long autocorrelations in physical quantities.
However, with a bin size of 250 sweeps, the values of $W(\vec(R),T)$ are 
decorrelated for all relevant $R$ and $T$\cite{stringbreak}.

\begin{figure}
\epsfig{figure=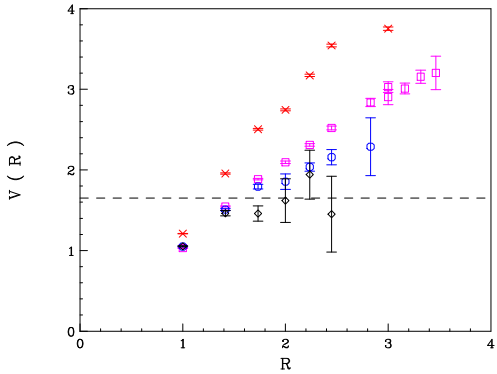, height=0.65\hsize}
\caption{Static energy measured using time slices: [0:1] (x's),
[1:2] (squares), [2:3] (circles) and [3:4] (diamonds). ($\kappa = 0.2050$)
The heavy-light pair threshold is denoted by the dashed line.}
\label{fig:static1}
\end{figure}

The results for the static energy ($\kappa = 0.2050$) are shown in 
Figure \ref{fig:static1}. 
Clear evidence for string breaking is seen for $T \ge 3$. 
This strengthens previous indications 
for string breaking\cite{stringbreak}.

We can also extract the Sommer scale 
($r_0^2 [V(r) - V(r-d)]/|d| = 1.65$) from
our determination of the static energy. Very roughly,
$r_0 \approx 1.25$ and $1/a \approx 500$ MeV.

\section{Spectrum of Light Hadrons}

Masses for the light hadrons can be determined 
for the various kappa values.
For $\kappa = .2050$ (highest statistics) we obtain:
$m_{\pi} = 0.391(6)$, $m_{\rho} = 1.008(8)$ and $m_P = 1.935(89)$.
Significant scale violation is seen with naive Wilson fermions.
For example, the scale $1/a$ determined from the proton mass is 485 MeV 
while the rho mass gives 764 MeV.

\begin{figure}
\epsfig{figure=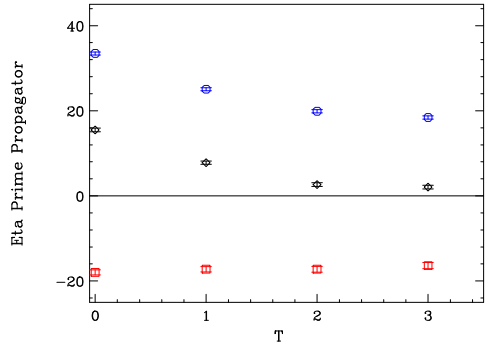, height=0.60\hsize}
\caption{The total $\eta '$ propagator for $\kappa = .2050$ 
on the $6^4$ lattices.
The valence term (circles), hairpins term (squares)
and total propagator (diamonds) are shown separately.}
\label{fig:etap}
\end{figure}

The relation between the axial $U(1)$ anomaly and the $\eta '$ mass 
is well understood in full QCD. 
For two light quarks ($N_f = 2$), 
$m^2_{\eta} = m^2_{\pi} + m^2_0$ where 
$m^2_0 = 2N_f\chi/f^2_{\pi}$ and the topological susceptibility is  
$V\chi \equiv <Q^2>_{\rm quenched}$.
The full $\eta$ propagator is the sum of a connected (valence quark) term and a
disconnected (hairpin) term.  Thus, in the continuum, the momentum space 
full propagator can be written: 
\begin{eqnarray}
 \lefteqn{(p^2 + m^2_{\pi} + m^2_0)^{-1}  =  ~~(p^2 + m^2_{\pi})^{-1}~-~ } 
               \\
  & & ~~~~~~~m^2_0(p^2 + m^2_{\pi})^{-1} 
             (p^2 + m^2_{\pi} + m^2_0)^{-1}   \nonumber 
\end{eqnarray}
These separate terms and their sum are shown in Figure \ref{fig:etap} for 
$\kappa = 0.2050$.
The cancellation between the valence and hairpin terms in the 
full propagator is evident.   
We obtain $m_{\eta'} = 1.16(12)$.
As expected the $\eta '$ mass is much heavier than the pion mass.


\section{Summary and Outlook}

Three results are clear from this study:
(1) The topological charge distribution behaves as expected in 
full QCD.
(2) Static energy shows clear evidence for string breaking for 
T $>$ 1 fm.
(3) Light hadron spectrum results on a $6^4$ coarse lattice look
reasonable. In particular, $m_{\eta '}a = 1.16(12) \gg m_{\pi}a = .391(6)$.

The full parametric studies needed
to optimize with TDA method will require
many runs on small sets of CPU's.
This might be done effectively by a 
seti@home approach \cite{phoenix}.

\vspace{-0.1in}


\end{document}